\documentclass[prd,aps,twocolumn,superscriptaddress,floatfix,notitlepage,nofootinbib,amssymb,amsmath]{revtex4-1}
\usepackage{mathtools}
\usepackage{epsfig}
\newcommand{\corr}[1]{#1}

\newcommand{\beqn}{\begin{eqnarray}}
\newcommand{\eeqn}{\end{eqnarray}}
\newcommand{\eq}[1]{(\ref{#1})}

\newcommand{\cL}{{\cal L}}

\newcommand{\Tr}{{\mathrm{Tr}\,}}

\newcommand{\Z}{{\mathbb Z}}
\newcommand{\bs}{\boldsymbol}

\newcommand{\avr}[1]{{\left\langle #1 \right\rangle}}

\newcommand{\R}{{\mathbb R}}
\newcommand{\T}{{\mathbb T}}
\newcommand{\bbS}{{\mathbb S}}

\begin{document}

\title{Phase structure of lattice Yang-Mills theory on ${\mathbb T}^2 \times {\mathbb R}^2$}

\author{M. N. Chernodub}
\affiliation{Institut Denis Poisson UMR 7013, Universit\'e de Tours, 37200 France}
\affiliation{Laboratory of Physics of Living Matter, Far Eastern Federal University, Sukhanova 8, Vladivostok, 690950, Russia}
\author{V. A. Goy}
\affiliation{Laboratory of Physics of Living Matter, Far Eastern Federal University, Sukhanova 8, Vladivostok, 690950, Russia}
\author{A. V. Molochkov}
\affiliation{Laboratory of Physics of Living Matter, Far Eastern Federal University, Sukhanova 8, Vladivostok, 690950, Russia}

\begin{abstract}
We study properties of SU(2) Yang-Mills theory on a four-dimensional Euclidean spacetime in which two directions are compactified into a finite two-dimensional torus ${\mathbb T}^2$ while two others constitute a large ${\mathbb R}^2$ subspace. This Euclidean ${\mathbb T}^2 \times {\mathbb R}^2$ manifold corresponds simultaneously to two systems in a (3+1) dimensional Minkowski spacetime: a zero-temperature theory with two compactified spatial dimensions and a finite-temperature theory with one compactified spatial dimension. Using numerical lattice simulations we show that the model exhibits two phase transitions related to the breaking of center symmetries along the compactified directions. We find that at zero temperature the transition lines cross each other and form the Greek letter $\gamma$ in the phase space parametrized by the lengths of two compactified spatial dimensions. There are four different phases. We also demonstrate that the compactification of only one spatial dimension enhances the confinement property and, consequently, increases the critical deconfinement temperature. 
\end{abstract}

\date{January 30, 2019}

\maketitle

\section{Introduction}

Lattice simulations indicate that Quantum Chromodynamics at zero density possesses a deconfining crossover transition at temperature $T_c \simeq 155\,\mathrm{MeV}$~\cite{ref:Fodor}. This important number sets a temperature scale, for example, in heavy-ion collisions. In a thermal equilibrium, a quantum field system, such as QCD, may be studied on a Euclidean space ${\mathbb R}^3 \times {\mathbb S}^1_T$ where the time coordinate is Wick-rotated into imaginary time which is further compactified into a circle of the length $L = 1/T$ related to temperature $T$.

In our article we discuss Yang-Mills theory on the Euclidean manifold ${\mathbb T}^2 \times {\mathbb R}^2$, where two (out of total four) Euclidean dimensions are compactified. The Euclidean ${\mathbb T}^2 \times {\mathbb R}^2$ theory may be treated as a theory at zero temperature with two compactified spatial dimensions. Alternatively, one may consider it as a finite-temperature theory with one compactified spatial dimension. In the latter interpretation, and for a small radius of compactified spatial dimension, a similar model, the deformed Yang-Mills theory, is expected to possess the Berezinskii-Kosterlitz-Thouless phase transition~\cite{ref:BKT} in a limit of a large number of colors~\cite{Simic:2010sv}. There are indications~\cite{Fraga:2016oul,Simic:2010sv} that the temperature of the phase transition should increase with the shrining radius of the compactification.  The phase diagram of a related  ${\mathbb T}^{2+D}$ theory with small radii of the $D$ dimensions has been studied in a large-$N$ limit with the use of a $1/D$ expansion in Ref.~\cite{Mandal:2011hb}. 

Properties of Yang-Mills theory at Euclidean ${\mathbb T}^2 \times {\mathbb R}^2$ space-time are also interesting because this system suffers from the  Linde problem~\cite{Linde:1980ts} which is much stronger compared to the one at finite temperature~${\mathbb S}^1 \times {\mathbb R}^3$~\cite{Fraga:2016oul}. In addition, the compactified spatial coordinates were suggested to affect the stabilization properties of the chromomagnetic vacuum~\cite{Elizalde:1996am}.

The associated equilibrium system would require geometrically constrained spatial dimensions. Influence of finite geometry of the space-time (such as spatial boundaries or spatial compactifications~\cite{Toms:1979ik}) on properties of quantum fields are usually associated with the Casimir effect. 

In (2+1) dimensions the non-Abelian Casimir effect in a spatial geometry bound by parallel ``chromometallic'' plates (wires) induces a smooth deconfining transition and leads to a new mass scale, the Casimir mass~\cite{Chernodub:2018pmt}. The Casimir mass is related to the mass of the magnetic gluon in (3+1)d Yang-Mills theory at finite-temperature~\cite{Karabali:2018ael}. The deconfining effect of the Casimir geometry and the effects of the mass gap are well understood in the confining compact Abelian gauge theory~\cite{ref:paper:1,ref:paper:2,ref:paper:3}.

In (3+1) dimensions the thermodynamics of the pure Yang-Mills theory in a finite box ${\mathbb T}^3 \times {\mathbb S}^1_T$ was studied numerically in Ref.~\cite{Bazavov:2007ny} where all spatial dimensions of the spatial torus ${\mathbb T}^3$ were kept of the same length. It turned out that for periodic boundary conditions the critical temperature $T$ decreases slowly as the volume of the spatial torus ${\mathbb T}^3$ is diminished. This property matches well with the asymptotic freedom of Yang-Mills theory: In a shrinking finite volume with periodic boundary conditions the lowest momentum of gluon rises and eventually reaches the perturbative scale where the gluons are weakly coupled so that the confinement is lost. On the contrary, in an open finite box with the confined (disordered) exterior of gluons the phase transition temperature shifts rapidly to higher values as the size of the box shrinks to zero~\cite{Bazavov:2007ny,Berg:2013jna}. The presence of the boundaries affects also the dynamics of fermions, as seen in effective infrared models of QCD~\cite{Ishikawa:1998uu,Tiburzi:2013vza,Flachi:2013bc,Flachi:2017cdo}. In our paper we concentrate on the pure gluonic phenomena and disregard possible effects of matter.

The interest in the Euclidean lattices with more than one compactified dimensions may also be understood in the context of the large-$N$ theories (in the planar limit). These studies were done both for $3+1$ and $2+1$ SU(N) gauge theories, both with asymmetrically compactified lattices possessing a sequence of the phase transitions~\cite{Bursa:2005tk} which were argued to match the structure of the phase transitions in the large-$N$ gauge theory at symmetric but finite-size ${\mathbb T}^4$ lattice at different values of the lattice coupling~\cite{Kiskis:2003rd}. The transitions are related to the patterns of the broken $\Z^4(N)$ center symmetry group associated with all four directions of the ${\mathbb T}^4$ torus. Contrary to a finite-$N$ gauge theory, in a large-$N$ limit finite volume effects disappear provided the size of the lattice exceeds certain critical value, as suggested in Ref.~\cite{Kiskis:2003rd} with further numerical evidence presented in Ref.~\cite{Narayanan:2004cp}. The related reviews may be found in Refs.~\cite{Teper:2004pk,Bursa:2005qm,Narayanan:2005en}. Our work may be regarded as a complement of the above large-$N$ studies with the focus on the smallest possible value, $N=2$.

The structure of our paper is as follows: in Section~\ref{sec:model} we briefly describe $SU(2)$ Yang-Mills theory that shares many properties of its QCD counterpart with three colors. We discuss the symmetries and the order parameters of the lattice theory. In Section~\ref{sec:phase} we present the numerical results on the phase diagram and space-time anisotropy of the gluon fields. We discuss two mentioned Minkowski realizations of the same Euclidean ${\mathbb T}^2 \times {\mathbb R}^2$ model (a zero temperature model with two compactified spatial dimensions and a finite-temperature model with one compactified spatial dimension). Our conclusions are summarized in the last section.

\section{Model and symmetries}
\label{sec:model}

\subsection{Lattice Yang-Mills theory on a torus}

The Yang-Mills (YM) theory is described by the following Lagrangian,
\beqn
\cL_{YM} = - \frac{1}{4} F_{\mu\nu}^a F^{\mu\nu,a},
\label{eq:L}
\eeqn
which is expressed via the field-strength tensor $F^a_{\mu\nu} = \partial_\mu A_\nu^a - \partial_\nu A_\mu^a + g f^{abc} A_\mu^b A_\nu^c$ of the non-Abelian (gluon) field $A_\mu^a$ with $a = 1, \dots N_c^2 -1$, and $f^{abc}$ are the structure constants of the $SU(N_c)$ gauge group related via $[T^a,T^b]= 2 i f^{abc} T^c$ to the generators ~$T^a$ of the $SU(N_c)$ group. In our paper we consider the simplest case of two colors, $N_c = 2$.

The lattice version of the $SU(2)$ Yang-Mills theory~\eq{eq:L} is given in terms of the $SU(2)$ matrices $U_l$ defined at the links $l \equiv \{x,\mu\}$ of the four-dimensional Euclidean cubic lattice with periodic boundary conditions in all four directions. The lattice matrix fields $U_l$ and the continuum vector fields ${\hat A}_\mu = T^a A_\mu^a$ are related as follows: 
\beqn
U_{x,\mu} = {\mathcal P} e^{ i g \int_{x}^{x+a\hat\mu} d x^\nu {\hat A}_\nu(x)}\simeq e^{i a g {\hat A}_\mu(x)},
\label{eq:U}
\eeqn
where ${\mathcal P}$ is the path--ordering operator and $a$ is the physical lattice spacing (equal to the length of the lattice links).

For the lattice version of the Yang-Mills theory~\eq{eq:L} we use the standard Wilson plaquette action:
\beqn
S_P = \beta \sum_P \left( 1 - \frac{1}{2}  \Tr U_P\right),
\label{eq:S}
\eeqn
where $U_{P_{x,\mu\nu}} = U_{x,\mu}U_{x+\hat\mu,\nu}U^\dagger_{x+\hat\nu,\mu} U^\dagger_{x,\nu}$ is the plaquette field strength and $\hat\mu$ is a unit lattice vector in the positive $\mu$ direction. The sum in Eq.~\eq{eq:S} is taken over all plaquettes of the lattice. For $N_c = 2$ gauge theory the lattice coupling constant $\beta$ is related to the continuum coupling $g$ as follows:
\beqn
\beta = \frac{4}{g^2}.
\label{eq:beta}
\eeqn

We perform our simulations at lattices of the geometries $N_1 \times N_2 \times N_s^2$, where the physical lengths\footnote{Notice that we use the notation $N_\mu$ for the lattice lengths expressed in lattice units and we reserve the notation $L_\mu$ notations for the same lengths expressed in physical units.}
\beqn
L_1 = a N_1, \qquad L_2 = a N_2,
\label{eq:LL}
\eeqn
determine the compactified directions of the torus ${\mathbb T}^2$  while the other two dimensions of the length $N_s$ correspond to the infinite plane ${\mathbb R}^2$. The lengths $N_1$ and $N_2$ of compactified torus ${\mathbb T}^2$ are always substantially smaller than the lengths $N_s$ of the plane ${\mathbb R}^2$. In an ideal case, the extension of the lattice in the ${\mathbb R}^2$ dimensions should be taken infinitely large, $N_s \to \infty$ (and we take $N_s \gg N_{1,2}$ in our work).

Technically, our simulations are completely analogous to the finite-temperature case which corresponds to the lattice geometry $N_t \times N_s^3$, where the physical length $L_t = a N_t$ of the single compactified dimension $N_t \ll N_s$ is related to the finite temperature in the standard way: $T = 1/L_t$. 

Physically, our simulations at the Euclidean lattice $N_1 \times N_2 \times N_s^2$ may be attributed to two different systems in Minkowski space-time. Firstly, the Euclidean lattice corresponds to a zero-temperature Yang-Mills theory defined in (3+1) dimensional spacetime, in which two of the three spatial dimensions are compactified into a torus with the lengths $L_1$ and $L_2$ given in Eq.~\eq{eq:LL}. As the third spatial direction remains infinite, the spatial three-dimensional manifold is ${\mathbb T}^2 \times {\mathbb R}$. Secondly, our simulations correspond to the Yang-Mills theory at a spatial manifold ${\mathbb S}^1 \times {\mathbb R}^2$ at a finite temperature $T = 1/L_1$, where the length of the compactified spatial dimension ${\mathbb S}^1$ is $L_2$ (obviously, with the permutation symmetry $L_1 \leftrightarrow L_2$).

\subsection{Order parameters}

\subsubsection{Finite temperature: one compactified dimension}

The Yang-Mills theory in (3+1) dimensional spacetime possesses a color confinement phase at low temperatures and a deconfinement phase at sufficiently high temperature. For the $SU$(2) gauge theory these phases are separated by a second order phase transition at the critical temperature~\cite{Fingberg:1992ju} 
\beqn
T_c  \equiv T_c^\infty = 0.69(2) \sqrt{\sigma}
\label{eq:Tc:infty}
\eeqn
where $\sigma$ is the tension of the string which confines color charges in fundamental representations (for instance, a test quark and a test antiquark) at zero temperature~$T{=}0$. The string tension $\sigma$ defines the physical mass and length scales in the theory. The superscript ``$\infty$'' in Eq.~\eq{eq:Tc:infty} indicates that the critical temperature $T_c^\infty$ corresponds to the thermodynamic limit of the system.

The order parameter of the deconfinement phase transition is the expectation value of the Polyakov line $P$ which is given by an ordered product of the non-Abelian matrices $U_l$ along the compactified (traditionally, $\mu=4$) ``temperature'' direction:
\beqn
P_{\bs x} = \frac{1}{2} \Tr \prod_{t=0}^{L-1} U_{{\bs x},x_4;4}\,,
\label{eq:PL}
\eeqn
where ${\bs x} \equiv (x_1,x_2,x_3)$ is the spatial three-dimensional coordinate. Equation~\eq{eq:PL} defines a gauge-invariant object because of the periodic boundary condition imposed along the compactified direction. The physical length of the compactified dimension is given by the inverse temperature, $L_t = 1/T$. 

The expectation value of the Polyakov line~\eq{eq:PL} determines the free energy $F$ of a single heavy quark, $\avr{P} = e^{-F/T}$. In the low-temperature phase the expectation value of the line is vanishing, $\avr{P} = 0$, indicating that the free energy of an isolated quark is infinite and quarks cannot exist as single objects. Thus, the low-temperature phase is confining: the quarks are confined inside hadrons that are colorless bound states of the quarks. In the deconfinement phase the Polyakov line acquires a nonzero value, $\avr{P} \neq 0$, implying the deconfinement property: the isolated quarks may exist.

The apparent existence of an order parameter implies also a presence of a symmetry which is spontaneously broken in one of the phases and unbroken in the other one. In the case of the deconfinement phase transition the relevant symmetry is represented by the center subgroup of the non-Abelian gauge group. Indeed, the lattice link fields~\eq{eq:U} transform under the gauge transformations $\Omega \in SU(N_c)$ as follows:
\beqn
U_{x\mu} \to U'_{x\mu} = \Omega_x U_{x\mu} \Omega_{x+\hat\mu}^\dagger.
\label{eq:Omega}
\eeqn
On a lattice with a compactified direction $\mu$ of the length $N_\mu$ the gauge transformations~\eq{eq:Omega} leave the gauge action~\eq{eq:S} invariant provided the gauge transformation matrix $\Omega_x$ is a quasi-periodic function of the compactified coordinate $x_\mu$:
\beqn
 \Omega_x \to  \Omega_{x+{\hat \mu} N_\mu} = C_\mu \Omega_x
\eeqn
where $C_\mu$ is an element of the center subgroup of the gauge group $SU(N_c)$ (the center subgroup is formed by all elements of the group of which commute with themselves and as well as with other elements of the group).

The Polyakov line~\eq{eq:PL} is sensitive to the transformations with respect to the center subgroup $\Z_{N_c}$ of the $SU(N_c)$ gauge group:
\beqn
P \to C_\mu P\,, \qquad C_\mu \in \Z_2.
\eeqn
In the confinement phase the expectation value of the line is vanishing, $\avr{P} = 0$, so that the center subgroup is unbroken. In the deconfinement phase the Polyakov line acquires a nonzero value, $\avr{P} \neq 0$, thus signaling the spontaneous breaking of the center symmetry.

\subsubsection{Two compactified dimensions}

In case of two compactified dimensions $x_1$ and $x_2$ it is natural to identify two Polyakov lines defined along the directions $\mu = 1,2$: 
\beqn
P^{(\mu)}_{\bs x} = \frac{1}{2} \Tr \prod_{t=0}^{L-1} U_{{\bs x},x_\mu;\mu},
\label{eq:L:mu}
\eeqn
where ${\bs x}$ corresponds to the remaining three coordinates.

The expectation values of the corresponding Polyakov lines are as follows:
\beqn
P^{(\mu)} = \frac{N_\mu}{N_1 N_2 N_s^2} \biggl\langle\biggl|\, \sum_{{\bs x}} P^{(\mu)}_{\bs x}\, \biggr| \biggr\rangle, 
\qquad
\mu = 1,2.
\label{eq:P:exp}
\eeqn

Given the topological isomorphism of the two-dimensional torus with the Cartesian product of two circles, ${\mathbb T}^2 \simeq  {\mathbb S}^1 \times {\mathbb S}^1$, one could suspect the existence of two phase transitions associated with breaking of two different $\Z_2$ symmetries that can be probed by the Polyakov lines along the corresponding compactified directions. In the next section we will check this hypothesis explicitly.

\section{Phase structure}
\label{sec:phase}

\subsection{Probing phase structure with Polyakov lines}

We studied the phase structure of  Yang-Mills theory at the lattices $N_1\times N_2\times N_s^2$ with the extensions $N_1, N_2 = 4, \dots 8$ in ${\mathbb T}^2$ torus dimensions and with the fixed $N_s = 32$ extension in the ``infinite'' ${\mathbb R}^2$ directions. For the finite-temperature case we also considered the additional lattice configuration with $N_1 = 4, \dots 8$ and $N_2=N_s=32$. We generated configurations of the lattice gauge fields using the Hybrid Monte Carlo algorithm~\cite{ref:Gattringer,ref:Omelyan}. We took 3 overrelaxation steps~\cite{ref:Gattringer} between trajectories in order to decrease an autocorrelation length between the configurations. We used $5\times 10^5$ trajectories for each sets of parameters. All observables were computed at every trajectory. Also we used binning for correct error estimation. We scanned the range of the lattice couplings $\beta\in [2.1,\,4.5]$ with the accuracy $\delta\beta=0.01$. All simulations were performed on Nvidia GPU cards.

\subsubsection{Expectations values of Polyakov lines}

\begin{figure}[!thb]
\begin{center}
\includegraphics[scale=0.5,clip=true]{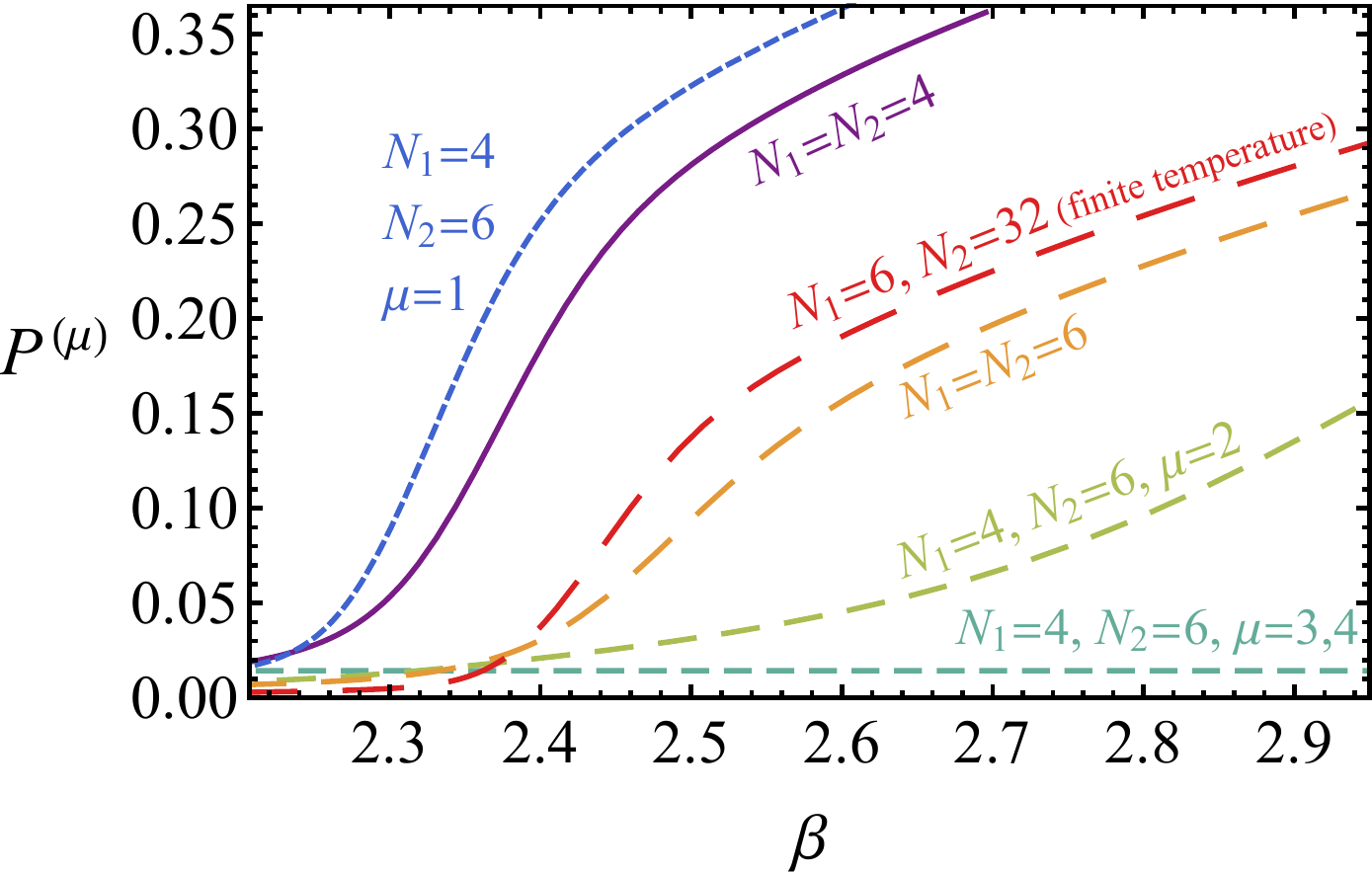}
\end{center}
\vskip -5mm 
\caption{Expectations values of Polyakov loops~\eq{eq:P:exp} in various directions at different lattice geometries (the detailed description is given in the text).}
\label{fig:lines}
\end{figure}

Typical examples of the expectation values of the Polyakov lines~\eq{eq:P:exp} at the ${\mathbb T}^2 \times {\mathbb R}^2$ geometry $N_1 \times N_2 \times N_s^2$ are shown in Fig.~\ref{fig:lines}. In the Figure we choose $N_1, N_2 = 4,6$ for two short compactified dimensions and $N_s = 32$ for two large (non-compactified in a thermodynamic limit) dimensions.

One may immediately notice from Fig.~\ref{fig:lines} that the Polyakov lines along the large (non-compactified) dimensions with $\mu=3,4$ are predictably insensitive to the compactification of other dimensions. On the contrary, each of the Polyakov lines closed along the compactified directions does experience of a transition-like behavior.

A common feature of the order parameters in the compactified directions is that they are small at low-$\beta$ region (hinting to an existence of a ``confinement-like'' phase at small $\beta$) and large in the high-$\beta$ limit (signaling a presence of the ``deconfinement-like'' region at large $\beta$). Of course, in our case of the spatial compactified dimensions these ``(de)con\-fi\-ne\-ment-like'' phases have nothing to do the real (de)confinement, and they should be characterized by the appropriate $\Z_2$ {\it spatial} symmetries.

Given our experience of the single-compactified ``temperature'' dimension, the very existence of these transitions is not surprising by itself. Moreover, it is natural that the transitions associated with the compact dimension of different lengths are taking place at different critical couplings~$\beta$, as it is indicated by the case of $N_1 = 4$, $N_2 = 6$ and $\mu=1,2$. What is interesting, however, is the interrelation of the transitions: Fig.~\ref{fig:lines} indicates that the compactification of one dimension influences the transition associated with the other compactified dimension. For example, the slope of the Polyakov line along the $N=6$ direction is shifting towards smaller $\beta$ as the other direction takes values $4,6$, and $32$.

\subsubsection{Susceptibilities of Polyakov lines and critical couplings}

\begin{figure}[!thb]
\begin{center}
\includegraphics[scale=0.5,clip=true]{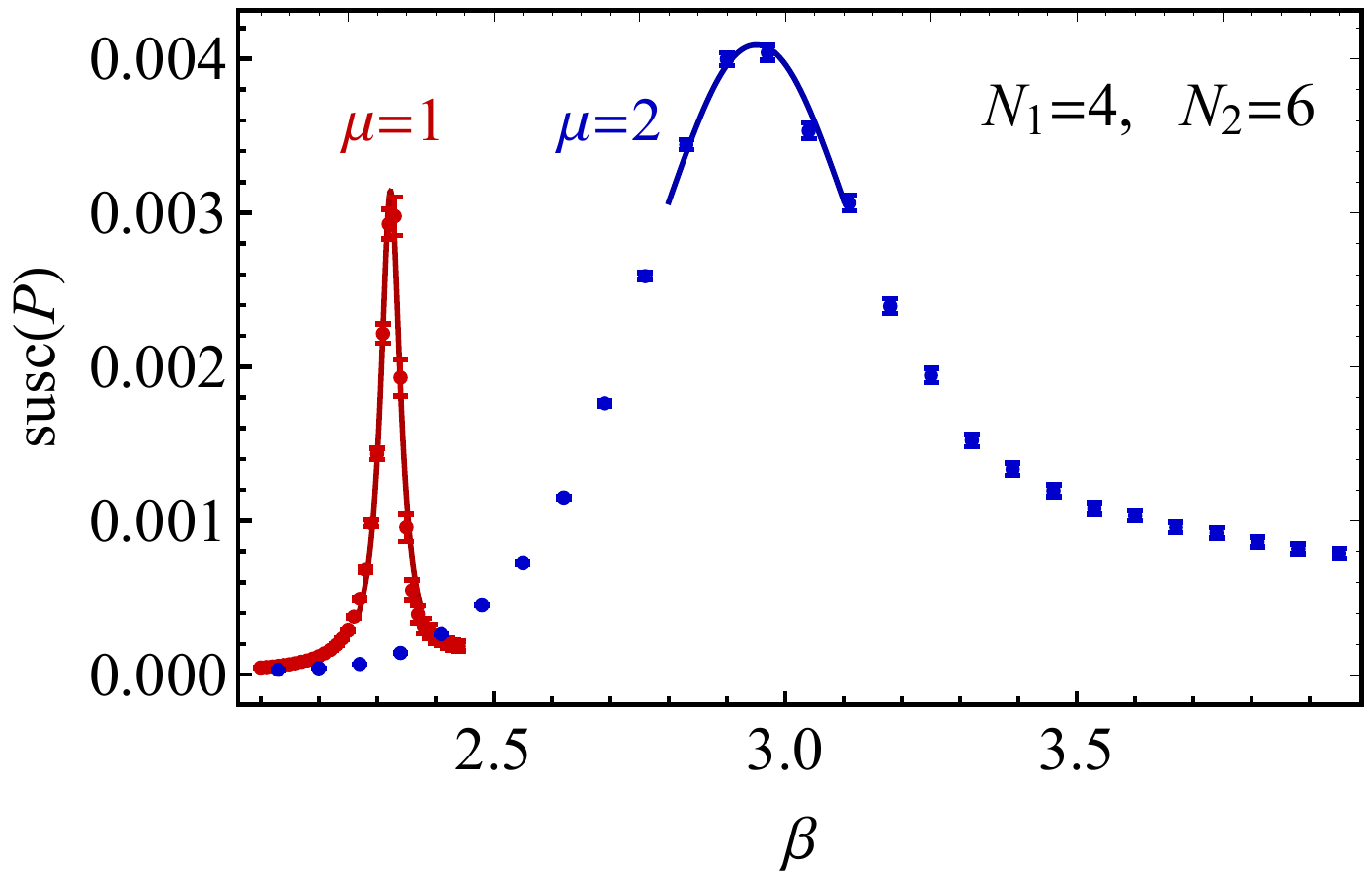}
\end{center}
\vskip -5mm 
\caption{Examples of the susceptibilities of the Polyakov lines~\eq{eq:susc:P} defined along the compactified directions $\mu = 1$ and $\mu = 2$ of unequal length ($N_1 = 4$ and $N_2 = 6$, respectively).}
\label{fig:susceptibilities}
\end{figure}

\begin{figure}[!thb]
\begin{center}
\includegraphics[scale=0.5,clip=true]{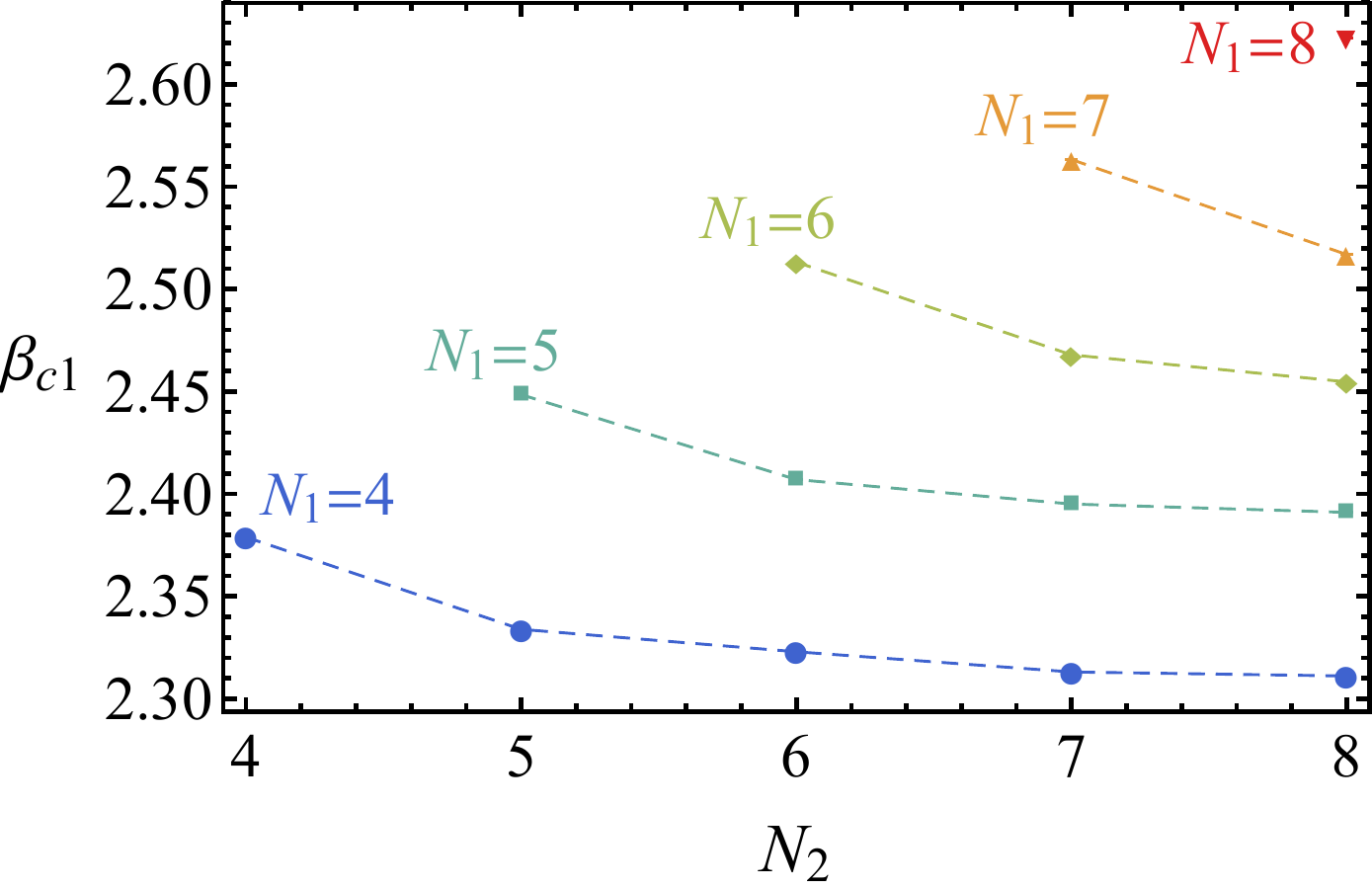}\\[3mm]
\includegraphics[scale=0.485,clip=true]{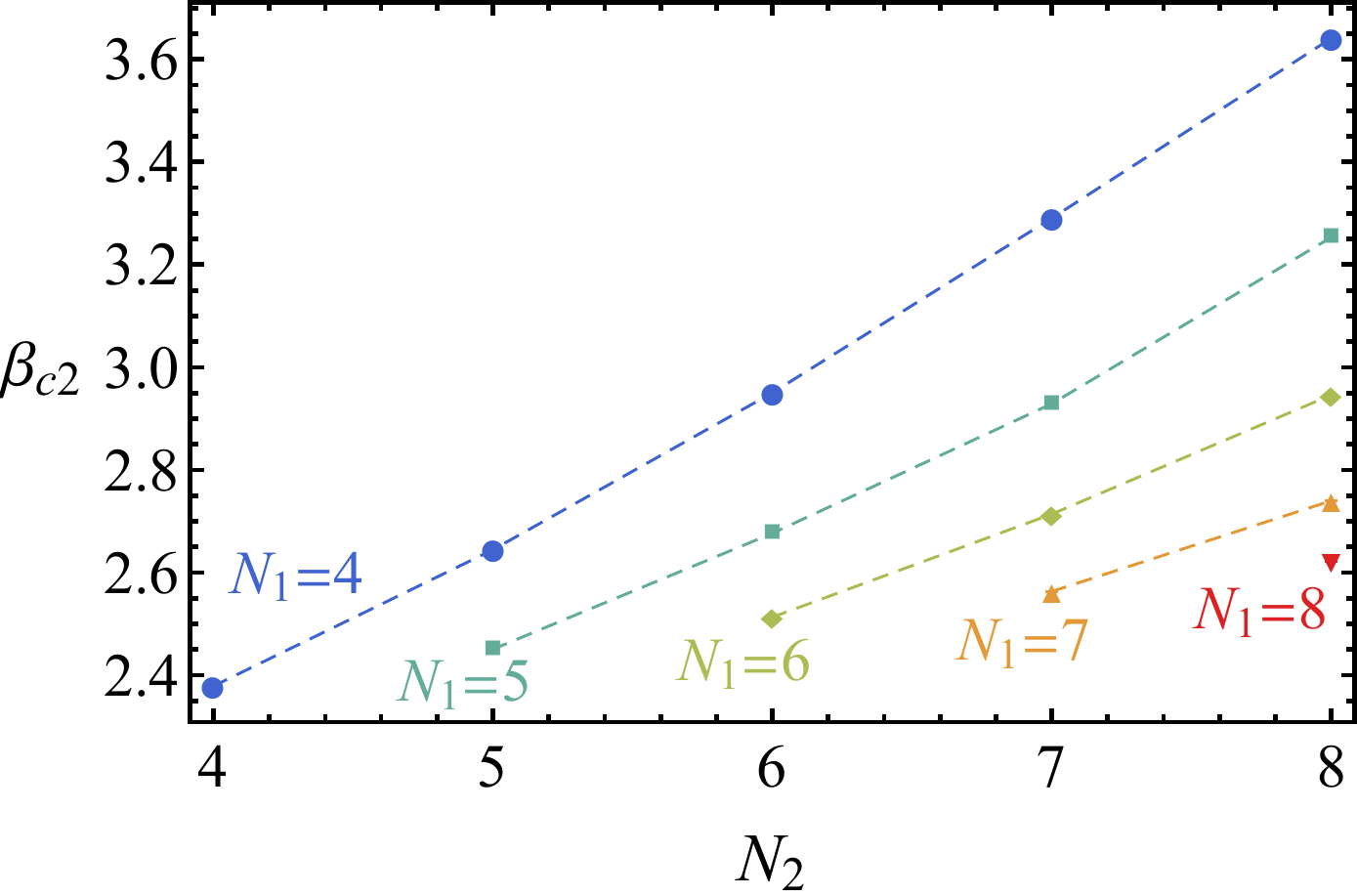}
\end{center}
\vskip -5mm 
\caption{The critical lattice couplings associated with $\mu=1$ (top) and $\mu=2$ (bottom) compactified dimensions on $N_1 \times N_2 \times 32^2$ lattice as a function of the lattice extension $N_2$ at the fixed sets of the lattice extension $N_1$ in the other direction. The dashed lines of are shown to guide an eye.}
\label{fig:betac}
\end{figure}

In a finite volume the points of a (pseudo) critical phase transition correspond to maxima of the susceptibility of the order parameter:
\beqn
{\mathrm{susc}}(P) = \frac{1}{\mathrm{Vol}} \left( \avr{|P|^2} - \avr{|P|}^2 \right).
\label{eq:susc:P}
\eeqn
In Fig.~\ref{fig:susceptibilities} we show an example of the susceptibilities of the Polyakov lines defined along the shorter ($\mu=1$) and longer ($\mu = 2$) compactified directions on the lattice with $N_1 = 4$, $N_2 = 6$ and $N_s = 32$. The two peaks appear at different coupling constant $\beta$ signaling the expected splitting of the phase transitions associated with breaking of the center symmetries along unequal directions. In Fig.~\ref{fig:betac} we show all pseudocritical lattice couplings~$\beta_{c\mu}$ for the lattices with $N_1,N_2 = 4, \dots 8$ and $N_s = 32$.

\subsubsection{Lattice spacing vs. lattice coupling}

\begin{figure}[!thb]
\begin{center}
\includegraphics[scale=0.5,clip=true]{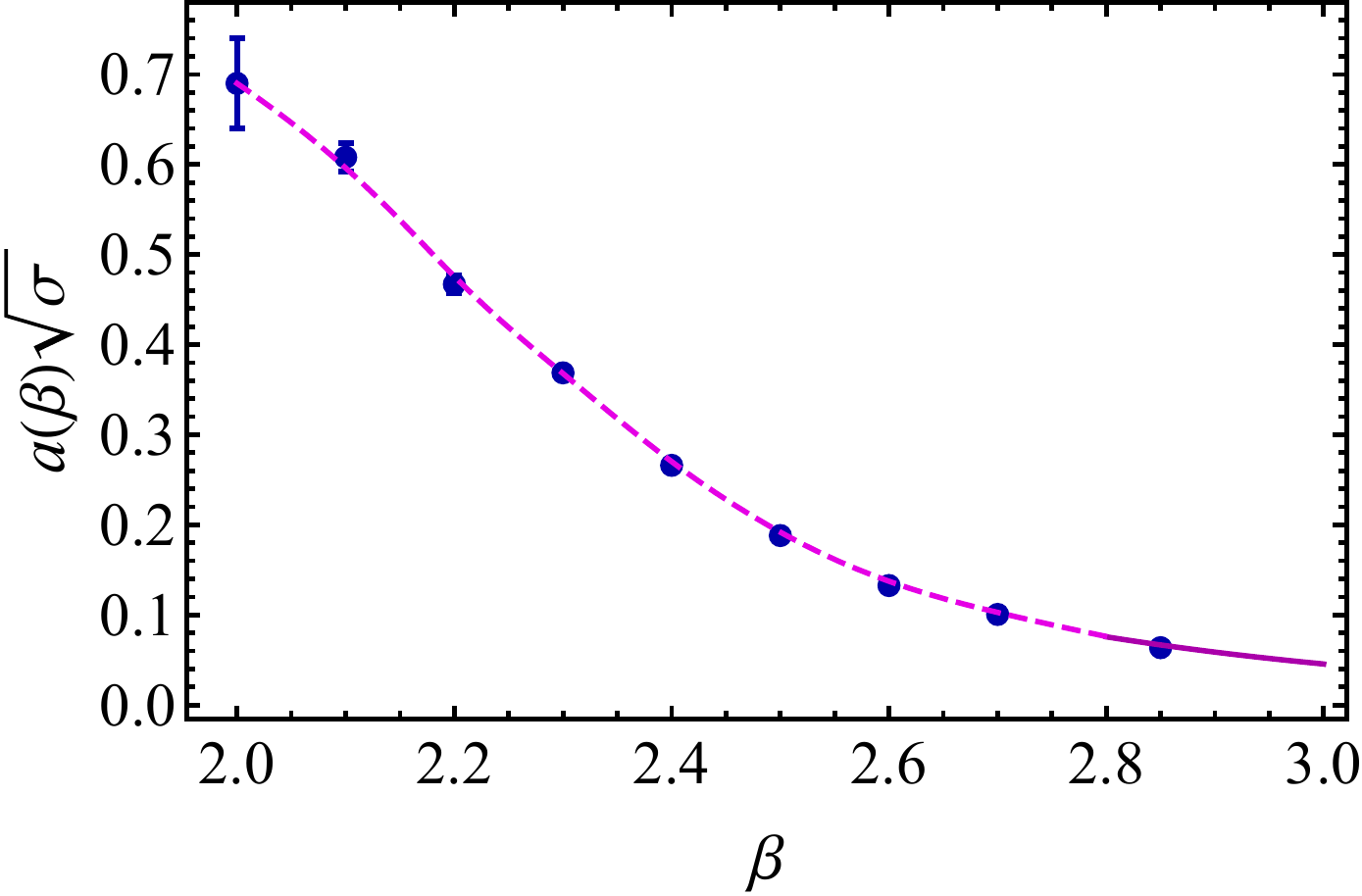}
\end{center}
\vskip -5mm 
\caption{Lattice spacing $a$ expressed via the physical string tension $\sigma$ and the lattice coupling $\beta$. The data points are taken from Ref.~\cite{Teper:1998kw}. The dashed line corresponds to a spline interpolation of the data, while the solid line is given by the two-loop renormalization group relation \eq{eq:a} and Eq.~\eq{eq:beta}.}
\label{fig:a:beta}
\end{figure}

The position of the pseudocritical lattice couplings, shown in Fig.~\ref{fig:betac}, correspond certain physical sizes (lengths $L_1$ and $L_2$) of the compactified dimensions~\eq{eq:LL}. In order to translate the lattice quantities to their continuum counterparts we need to know the physical value of the lattice spacing $a$ as the function of the lattice coupling~$\beta$. To this end we follow Ref.~\cite{Teper:1998kw}: in the low-$\beta$ region (at strong and moderate coupling) we interpolate the data for the lattice string tension, obtained in Ref.~\cite{Teper:1998kw} as well, using a spline function. In the scaling window at the weak coupling ($g \ll 1$ or $\beta \gg 1$) the dependence of the lattice spacing $a$ on the $SU(2)$ coupling constant $g$ may be determined by the renormalization group equation. In two loops (see, e.g., Ref.~\cite{Fingberg:1992ju}):
\beqn
a\left( g^2 \right) \Lambda_L = \exp\left\{ - \frac{12 \pi^2}{11 g^2} + \frac{51}{121} \ln \frac{24 \pi^2}{11 g^2} \right\},
\label{eq:a}
\eeqn
where $\Lambda_L$ is a mass scale. The two approaches are matched at $\beta = 2.8$, and the result for the lattice spacing is shown in Fig.~\ref{fig:a:beta}.

\subsection{Zero-temperature SU(2) gauge theory\newline with two compactified spatial dimensions}

In this section we consider the interpretation of our Euclidean results in terms of a zero-temperature $SU(2)$ gauge theory in Minkowski spacetime with two compactified spatial dimensions. 

\subsubsection{Phases of $SU(2)$ YM at ${\mathbb T}^2 {\times} {\mathbb R}$ spatial manifold at $T{=}0$}

\begin{figure}[!thb]
\begin{center}
\includegraphics[scale=0.475,clip=true]{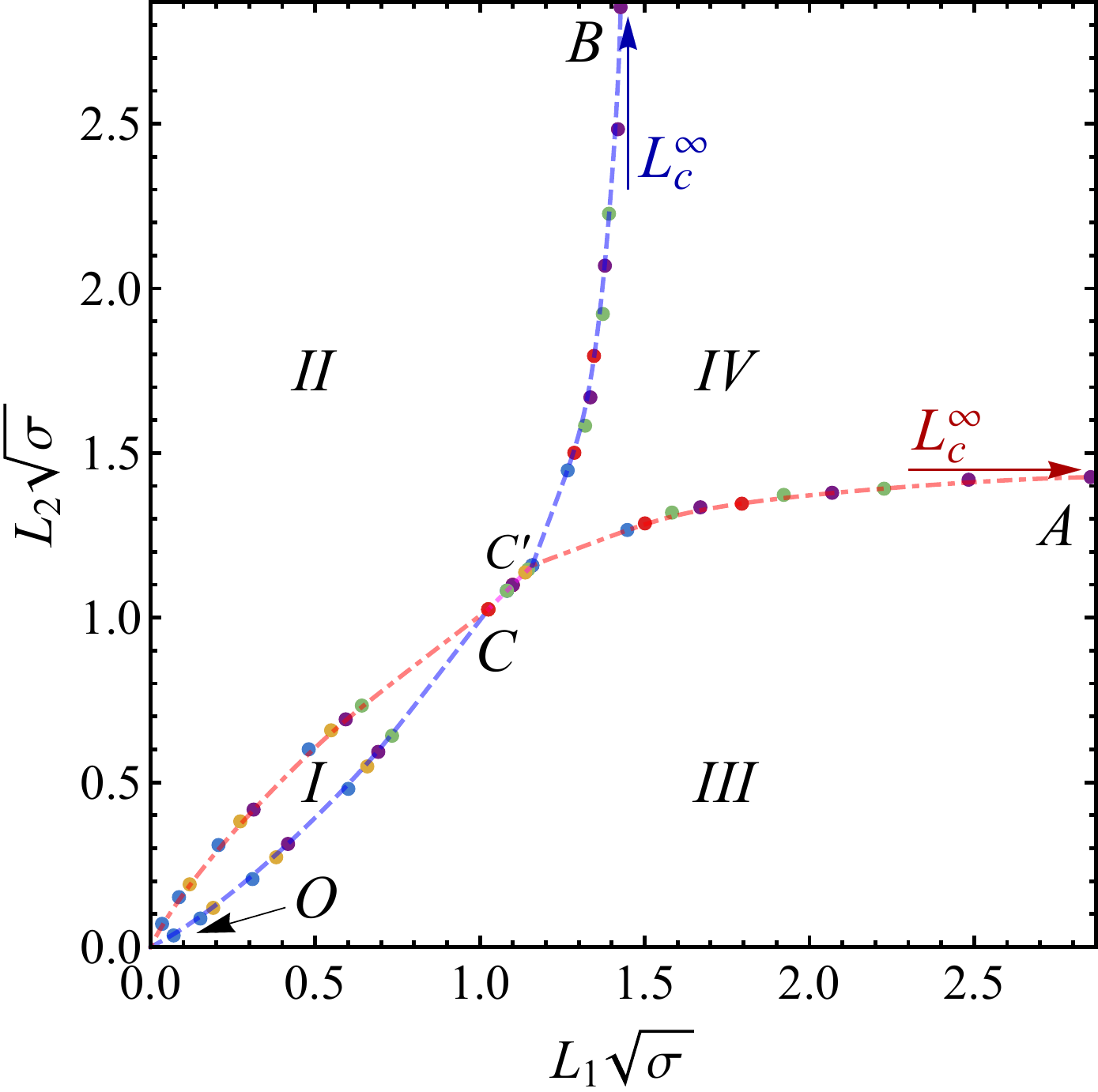}
\end{center}
\vskip -5mm 
\caption{The phase diagram of the zero-temperature SU(2) Yang-Mills theory with two compactified spatial dimensions $L_1$ and $L_2$. The colored arrows point to the asymptotic critical lengths~\eq{eq:Lc} of compactification transition related to a finite-temperature phase transition~\eq{eq:Tc:infty}. The colored points correspond to the numerical data, the dashed blue line (the dot-dashed red line) is the $L_1$ ($L_2$) compactification transition. A detailed description is given in the text.}
\label{fig:phase:diagram}
\end{figure}

Using the results for the critical couplings $\beta_{c1}$ and $\beta_{c2}$ (given in Fig.~\ref{fig:betac}) and the dependence of the lattice spacing~$a$ on the lattice coupling $\beta$ (presented in Fig.~\ref{fig:a:beta}) we may reconstruct the phase diagram of the model in terms of the physical lengths of the compactified dimensions $L_1$ and $L_2$, Eq.~\eq{eq:LL}. The phase diagram is shown in Fig.~\ref{fig:phase:diagram}. 

In the phase diagram there are two critical lines corresponding to the breaking of the center symmetry in each of the compactified dimensions. Both these lines start at the point $O$ at the origin, $L_1 = L_2 = 0$, then cross each other at the point $C$, follow a common line till the point $C'$ and then deviate from each other in the appropriate limits $L_{1,2} \to \infty$. Notice that in each of these limits, either at $L_1 \to \infty$ or at $L_2 \to \infty$, the Euclidean ${\mathbb T}^2 \times {\mathbb R}^2$ space reduces to the finite-temperature case ${\mathbb S}^1 \times {\mathbb R}^3$ for which the point of the phase transition is well known~\eq{eq:Tc:infty}. Given the identification of the length of the single compactified dimension with temperature, $L = 1/T$, one gets the compactification transitions at $L_2 = L_c^\infty$ (point $A$ in Fig.~\ref{fig:phase:diagram}) and $L_1 = L_c^\infty$ (point $B$), respectively. Here the critical length of the ``compactification'' transition is given by Eq.~\eq{eq:Tc:infty}:
\beqn
L_c^\infty = \frac{1}{T_c^\infty} = 1.45(4) \frac{1}{\sqrt{\sigma}}.
\label{eq:Lc}
\eeqn

In Fig.~\ref{fig:phase:diagram} the transition lines resemble the Greek letter~$\gamma$. The phase diagram possesses four phases marked by the Roman numerals in Fig.~\ref{fig:phase:diagram}. These phases are characterized by different patterns of spontaneous breaking of spatial center symmetries associated with the two compactified dimensions:

\begin{itemize}

\item[I.] $P^{(1)} \neq 0$ and $P^{(2)} \neq 0$ (both spatial center symmetries $\Z_2$ are broken spontaneously); 

\item[II.] $P^{(1)} \neq 0$ and $P^{(2)} = 0$ (one $\Z_2^{(1)}$ is broken while the other $\Z_2^{(2)}$ is unbroken); 

\item[III.] $P^{(1)} {=} 0$ and $P^{(2)} \neq 0$ (unbroken $\Z_2^{(1)}$, broken $\Z_2^{(2)}$); 

\item[IV.] $P^{(1)} = 0$ and $P^{(2)} = 0$ (both spatial center symmetries $\Z_2$ are unbroken). 

\end{itemize}

We expect that in the continuum limit the line of the common phase transition $C-C'$ should shrink into a single point since all the data points at this line are represented by the lattices with equal compactified lengths $N_1=N_2$. The length of the $C-C'$ segment gives the order of our systematic accuracy due to finite-volume effects at $L_1 \sim L_2 \sim 1/\sqrt{\sigma}$.

The phase diagram~\eq{fig:phase:diagram} of the $SU(2)$ gauge theory qualitatively agrees with the diagram of a $SU(N)$ Yang-Mills theory in the limit of the large number of colors $N \to \infty$ on the ${\mathbb T}^{2+D}$ torus obtained in Ref.~\cite{Mandal:2011hb} with the use of a $1/D$ expansion assuming that the radii of the $D=2$ dimensions are small.

Our results indicate that compactification of one or two spatial dimensions does not induce a real deconfinement transition. Indeed, we consider the theory at zero temperature while the Polyakov lines along any of the non-compactified dimension are effectively zero. The two discussed transitions break only the spatial $\Z_2$ symmetries while the temporal $\Z_2$ symmetry, which is responsible for the color confinement phenomenon, remans unbroken. The system always resides in the color confinement phase.

\subsubsection{Anisotropy in gluon fields}

One may expect that the fluctuations of the gluon fields should be affected by the anisotropic nature of the spacetime~\cite{Elizalde:1996am}. In order to explore the anisotropic in the gluon fluctuations we calculate the following gauge invariant quantity (no sum over the indices $\mu$ and $\nu$ is implied):
\beqn
\delta \avr{F_{\mu\nu}^2} = \avr{F_{\mu\nu}^2} - \frac{1}{12} \sum_{\alpha,\beta=1}^4 \avr{F_{\alpha\beta}^2}.
\label{eq:delta:F}
\eeqn
This quantity compares the field-strength fluctuations $\avr{F_{\mu\nu}^2}$ (with fixed $\mu$ and $\nu$) to the mean fluctuations (averaged over all possible geometrical orientations). An additive quartically-divergent ultraviolet contribution to the field-strength tensor squared drops out in the difference~\eq{eq:delta:F}. Below we call the quantity~\eq{eq:delta:F} ``the gluon anisotropy'' for shortness.

\begin{figure}[!thb]
\begin{center}
\includegraphics[scale=0.7,clip=true]{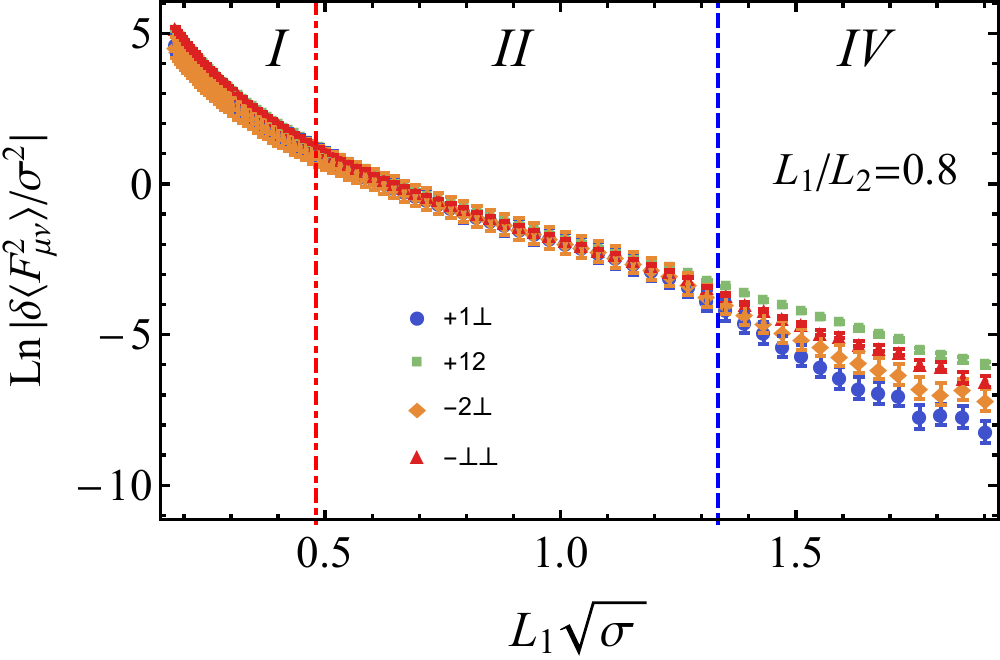}\\[3mm]
\includegraphics[scale=0.7,clip=true]{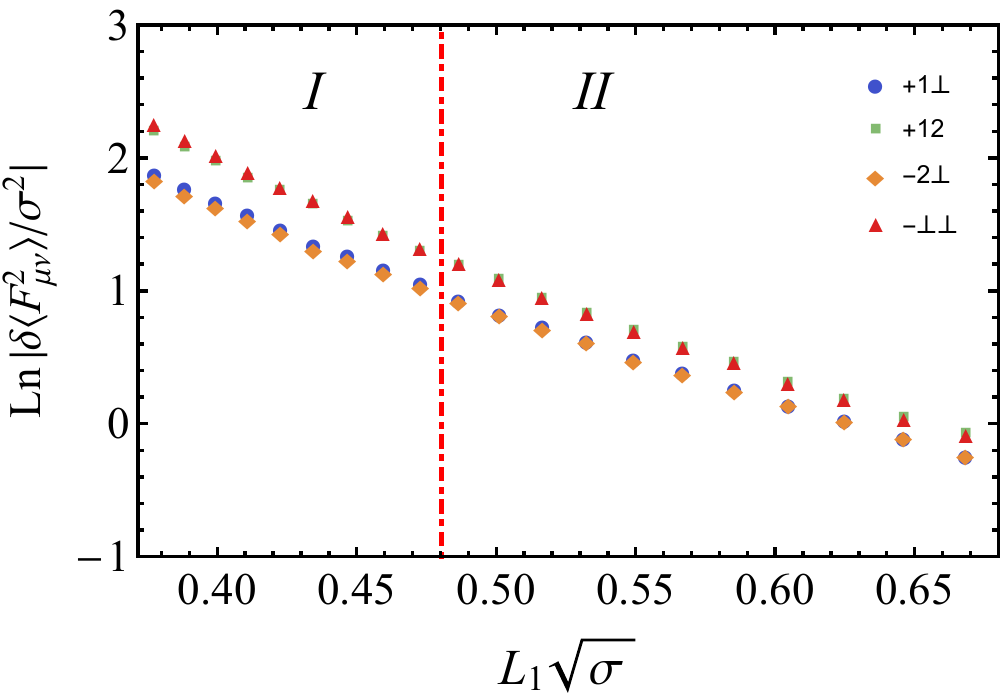}
\end{center}
\vskip -5mm 
\caption{The natural logarithm of the gluon anisotropy~\eq{eq:delta:F} vs. the length of the compactified direction $L_1$ for fixed ratios between the compactification lengths~$L_1/L_2 = 0.8$. All three representative phases are shown. The notation for the labels ``$\pm\mu\nu$'' is as follows:  ``$\pm$'' implies a positive/negative value of the anisotropy~$\delta F^2_{\mu\nu}$ and $\mu = \perp$ denotes the long direction, either $\mu=3$ or $\mu=4$. The lower figure is a zoom in on the I-II transition of the whole plot (the upper figure). The vertical dot-dashed red and dashed blue lines mark the positions of the phase transition lines of Fig.~\ref{fig:phase:diagram}.}
\label{fig:anisotropy}
\end{figure}

In Fig.~\ref{fig:anisotropy} we show the gluon anisotropy~\eq{eq:delta:F} along the straight path in the phase space of Fig.~\ref{fig:phase:diagram} which keeps the ratio $L_1/L_2 = 0.8$ constant. This path first crosses the transition corresponding the longer compactification length $L_2$ (the dot-dashed red line) and then it goes through the transition corresponding to the shorter compactification length $L_1$ (the dashed blue line).

Let us consider the proportionally increasing lengths of the compactified dimensions $L_1$ and $L_2$, with $L_1/L_2 = 0.8$ fixed, Fig.~\ref{fig:anisotropy}.  At sufficiently small dimensions of the compactified directions we are in the phase ``I'', where both spatial $\Z_2$ symmetries are broken, and the hierarchy of the gluon asymmetries is well visible in Fig.~\ref{fig:anisotropy}(bottom): the gluon anisotropy in the compactified $\T^2$  and non-compactified $\R^2$ dimensions are, respectively, positive- and negative- valued, and equal in the absolute values. The mixed ${\mathbb S}^1 \times \R$ are positive- and negative- valued for ${\mathbb S}^1$ corresponding to the shorter ($L_1$) and longer ($L_2$) length, respectively (and they both have the same absolute value as well). Thus, (the absolute value of) the gluon anisotropy has a double-degenerate structure.

With increasing proportionally the size of the compactified dimensions we cross the first transition point between the phases ``I''-``II'', where $\Z_2$ of the longer compactification $L_2$ gets restored. According to Fig.~\ref{fig:anisotropy} this transition does not seem to affect the gluon asymmetries. 

As we increase the compactified length even further, we move along the phase ``II'', approach the transition line ``II''-``IV'' and then cross to the phase ``IV'' where both $\Z_2$ symmetries are restored. At the second transition point the mentioned double degeneracy disappears in a continuous manner: all four gluon asymmetries possess different strengths in the vicinity of the second transition and in the whole ``IV'' phase.

\corr{
The series of transitions on ${\mathbb T}^3$ tori were discussed earlier in the context of the large but finite-$N$ gauge theories on very asymmetric lattices in Ref.~\cite{Bursa:2005tk}. Interestingly, these transition were conjectured in Ref.~\cite{Bursa:2005tk} to be related to the series of the finite-volume infinite-$N$ phase transitions on symmetric ${\mathbb T}^4$ lattices, as conjectured in Ref.~\cite{Kiskis:2003rd} with further numerical support obtained in Ref.~\cite{Narayanan:2004cp}. Here we concentrate on the phase diagram of the $SU(N)$ gauge with the smallest possible value $N=2$ on the torus ${\mathbb T}^4$ with two compactified dimensions and two non-compactified (formally, infinite-lengths, ${\mathbb T}^2 \to {\mathbb R}^2$) dimensions.}

\subsection{Finite-temperature SU(2) gauge theory\newline with one compactified spatial dimension}

\subsubsection{Phases of $SU(2)$ YM at ${\mathbb S}^1 {\times} {\mathbb R}^2$ spatial manifold at $T{>}0$}

\begin{figure}[!thb]
\begin{center}
\includegraphics[scale=0.25,clip=true]{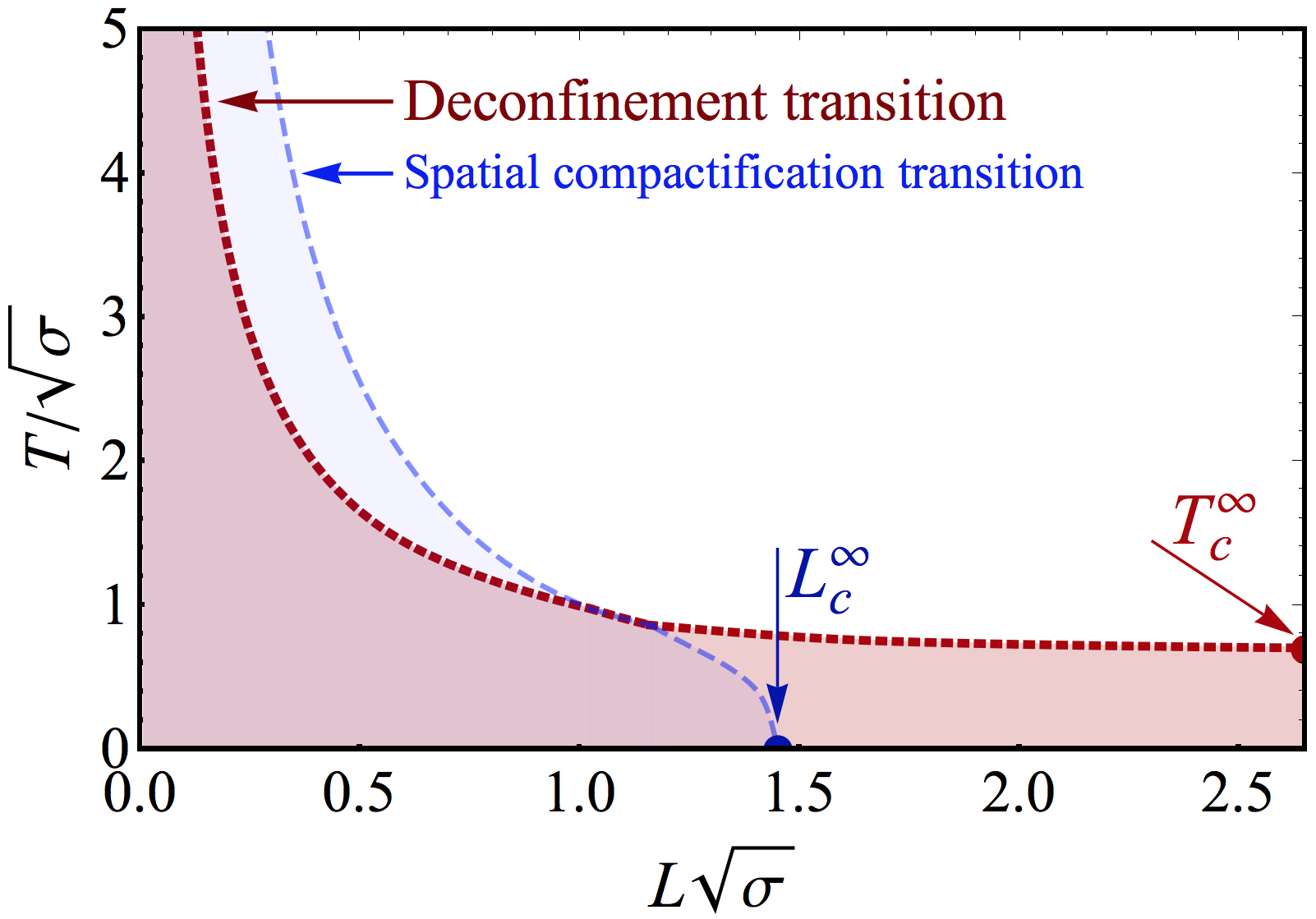}
\end{center}
\vskip -5mm 
\caption{The phase diagram of finite-temperature $SU(2)$ Yang-Mills theory with one compactified spatial dimension of the length $L$. The thick red line is the deconfinement phase transition and the thin blue line represents the critical line of the breaking of the global $\Z_2$ symmetry associated with the compactified spatial dimension. The red- and blue- shaded areas are the confinement and spatial--$\Z_2$ restored phases, respectively. The critical temperature~\eq{eq:Tc:infty} and the critical compactified length~\eq{eq:Lc} are shown by the arrows.}
\label{fig:phase:diagram:T}
\end{figure}

Now let us turn to the second interpretation of our numerical results by considering the finite-temperature $SU(2)$ Yang-Mills model with one compactified spatial dimension. Taking the phase diagram of Fig.~\ref{fig:phase:diagram} as a starting point and setting $L_1 = L$ as the size of the single compactified spatial dimension and $L_2 = 1/T$ as the imaginary time, we get the phase diagram shown in Fig.~\ref{fig:phase:diagram:T}. For the sake of clearness we omitted the numerical data points in this figure and we show the spline interpolations only. The agreement of the numerical data and the interpolations is as good as in Fig.~\ref{fig:phase:diagram}.

Figure~\ref{fig:phase:diagram:T} shows that the compactification of one of the spatial dimensions increases the deconfinement temperature. This observation may indicate that the compactification of only one dimension enhances the confinement property of the vacuum as one needs stronger thermal fluctuations to destroy the color confinement. 

At large compactified dimensions $L$ the deconfinement temperature is almost independent of $L$ and, obviously, $T_c(L) \to T_c^\infty$ as $L \to \infty$. 

As the compactified length decreases to the critical length $L = L_c^\infty$, given in Eq.~\eq{eq:Lc}, the spatial $\Z_2$ gets broken at zero temperature. At the point $(L_s,T_s) \simeq (1.1/\sqrt{\sigma},0.9\sqrt{\sigma})$ the two transitions lines cross and the confinement region (with unbroken temporal $\Z_2$ symmetry) is always accompanied by the breaking of the spatial $\Z_2$ symmetry. Then the diminishing length of the compactified dimension $L$ leads to a rapid increase of the deconfinement temperature, in agreement with analytical arguments of Refs.~\cite{Simic:2010sv,Fraga:2016oul}. Thus, we conclude that the breaking of the spatial $\Z_2$ symmetry of the compactified direction enhances the stability of the color confinement against the thermal fluctuations. In other words, the confinement of color is enhanced by compactification of one of the spatial dimensions into a periodic circle.

\subsubsection{Non-Abelian Casimir effect and magnetic sector of QCD}

The enhancement of the critical deconfinement temperature with the compactification of one of the spatial coordinates may be treated as a nonperturbative feature of a non-Abelian version of the Casimir effect. We would like to stress that the enhancement of the critical temperature comes as a result of the restricted Casimir-type geometry and not as a finite-volume effect, as the total volume is still large (and, moreover, we are formally close to the infinite volume limit).

Consider a finite-temperature $3+1$ dimensional Yang-Mills theory with one compactified spatial direction. The corresponding manifold is $\bbS^1_S \times \bbS^1_T \times \R^2$, where $\bbS^1_S$ corresponds to the compactified spatial direction and $\bbS^1_T$ is the imaginary time (temperature) dimension. Let us take a limit where the compactified spatial direction tends to zero, $L \to 0$, so that the spatial sub-manifold $\bbS^1_S$ gradually shrinks to a point. We may now treat the shrinking spatial dimension as if it is a compactified imaginary time dimension in the limit of infinitely high temperatures. We immediately arrive to the picture of the dimensional reduction, in which the gluon components corresponding to the compactified directions become infinitely heavy and we are left with a finite-temperature $2+1$ Yang-Mills theory at the three-dimensional manifold $\bbS^1_T \times \R^2$. The gauge coupling $g_3$ of the resulting three dimensional theory is a quantity of the dimension of mass${}^{1/2}$:
\beqn
g_3(L) = \frac{g}{\sqrt{L}}\,,
\label{eq:g3}
\eeqn
where $L$ plays a role of the new ``inverse temperature $1/T$''. Indeed, if the small spatial dimension would be a finite-temperature direction when we would call the quantity~\eq{eq:g3} as the ``magnetic constant'' (with the appropriate redefinition $L \to 1/T$) which would then set up a non-perturbative scale for the magnetic gluons. The corresponding magnetic mass would be proportional to the three-dimensional coupling~\eq{eq:g3} and it would give a mass scale for the spatial (magnetic) string tension~$\sigma_3$ in (2+1) dimensional SU(2) gauge theory.

The first-principle numerical simulations indicate that the critical temperature of the deconfining phase transition in (2+1) dimensions is related to the (2+1)-{\it{dimensional}} string tension $\sigma_3$ as follows~\cite{Teper:1993gp}:
\beqn
T_{3,c} = C_T \sqrt{\sigma_3}\,, \qquad C_T = 1.121(8)\,,
\label{eq:Michael:T}
\eeqn
while the string tension is proportional to the magnetic coupling squared~\cite{Teper:1998te}:
\beqn
\sqrt{\sigma_3} = C_\sigma g^2_3\,, \qquad C_\sigma = 0.3353(18)\,.
\label{eq:Michael:sigma}
\eeqn
These two relations, along with Eq.~\eq{eq:g3}, give us the following prediction for the critical temperature in (3+1)d Yang-Mills theory with one compactified spatial coordinate of sufficiently short length $L$:
\beqn
T_{{\mathbb S}^1 \times \R^2}(L) = C_{{\mathbb S}^1 \times \R^2} \frac{g^2}{L}, \qquad L \to 0\qquad
\label{eq:Tc:L0}
\eeqn
where $g$ is the coupling of the original (3+1)d Yang-Mills theory and 
\beqn
C^{\mathrm{th}}_{{\mathbb S}^1 \times \R^2} \equiv C_T C_\sigma = 0.376(9),
\label{eq:C:0:th}
\eeqn
is a phenomenological dimensionless constant which is determined by the critical temperature of the (2+1) Yang-Mills theory, Eqs.~\eq{eq:Michael:T} and \eq{eq:Michael:sigma}.

Our numerical data of Fig.~\ref{fig:phase:diagram:T} indicate that the asymptotic relation~\eq{eq:Tc:L0} indeed works with $C^{\mathrm{num}}_{{{\mathbb S}^1 \times \R^2}} = 0.44(4)$ which is reasonably close to the estimation in Eq.~\eq{eq:C:0:th}. The small difference between these quantities may be attributed to systematic numerical errors  in the limit $L \to 0$. A careful treatment this small discrepancy would require significant numerical simulations, and therefore we leave it beyond the scopes of the present paper.

We also conclude that in the limit of a small compactification $L \to \infty$ of the spatial dimension the phase transition in the $SU(2)$ gauge theory should be of the second order with the Ising type of the universality class similarly to the case of the $SU(2)$ lattice gauge theory in (2+1) dimensions~\cite{Teper:1993gp}. In the large $N_c$ limit the phase transition of Berezinskii-Kosterlitz-Thouless~\cite{ref:BKT} is expected~\cite{Simic:2010sv}.

\section{Conclusions}

In our paper we study, using first-principle numerical simulations, a $SU(2)$ Yang-Mills theory in a four-dimensional Euclidean spacetime with two compactified spatial dimensions of unequal lengths. The other two dimensions were chosen to be sufficiently large, so that the geometry of our space-time corresponds to a direct product of a two-dimensional torus and flat two-dimensional space ${\mathbb T}^2 \times {\mathbb R}^2$. In Minkowski space-time this geometry may be associated to two different systems. We may interpret it either as a zero-temperature theory in the three-dimensional manifold with two compactified spatial dimensions ${\mathbb T}^2 \times {\mathbb R}$ or as a finite-temperature theory on ${\mathbb T}^2 \times {\mathbb R}$ spatial manifold with one compactified dimension. Our work may be regarded as an $N=2$ complement of the studies of the $SU(N)$ gauge theories in large-$N$ limit on symmetrically~\cite{Kiskis:2003rd,Narayanan:2004cp} and asymmetrically~\cite{Bursa:2005tk} compactified lattices featuring sequence of the finite-size-free phase transitions in a finite volume.

In our work, we show that the phase diagram of the zero-temperature Yang-Mills theory on a ${\mathbb T}^2 \times {\mathbb R}$ spatial manifold possesses two critical transitions associated with breaking of two different center ($\Z_2 \times \Z_2$) spatial symmetries. The compactifications ${\mathbb R}^2 \to {\mathbb S}^1 \times {\mathbb S}^1 \simeq {\mathbb T}^2$ lead to the breaking of the appropriate $\Z_2$ symmetries, which are probed with the help of the spatial Polyakov lines. Both transitions affect each other so that the resulting phase diagram, parametrized by the lengths of the compactified dimensions, resembles the Greek letter $\gamma$, as shown in Fig.~\ref{fig:phase:diagram}. In addition, our results have revealed that the compactification of two out of three spatial dimensions does not lead to a color deconfinement at zero temperature.

These results also reveal the phase diagram of a finite-temperature Yang-Mills theory on a spatial manifold ${\mathbb S}^1 \times {\mathbb R}^2$ in which only one spatial dimension is compactified to a circle of the length $L$ while two other dimensions remain (infinitely) large. The system possesses two phase transitions in the $T{-}L$ plane: one transition line is a true deconfinement phase transition while the second line is associated with the breaking of the spatial $\Z_2$ symmetry, Fig.~\ref{fig:phase:diagram:T}. These two transition lines form the Greek letter $\chi$. We see that the compactification of one spatial dimension increases the critical deconfinement temperature thus enhancing the color-confinement property. We argue that the enhancement of the critical temperature is not a finite-volume effect and it comes as a result of the restricted Casimir-type spatial geometry of an infinite-volume system.

\begin{acknowledgments}

The authors thank Emilio Elizalde for e-mail communications which inspired this work. We are grateful to Eduardo Fraga, Daniel Nogradi and Jorge Noronha for comments. We thank Takeshi Morita and Gautam Mandal for making us aware about their work~\cite{Mandal:2011hb}. The numerical simulations were performed at the computing cluster Vostok-1 of Far Eastern Federal University. The research was carried out within the state assignment of the Ministry of Science and Education of Russia (Grant No. 3.6261.2017/8.9). 

\end{acknowledgments}

\end{document}